\documentclass[twocolumn]{revtex4}
\pdfoutput=1 

\usepackage[T1]{fontenc} 
\usepackage[english]{babel}

\usepackage{graphicx}
\usepackage{amsmath,amsthm,amssymb,amsfonts}
\usepackage{mathrsfs,bbm,slashed}
\usepackage{hyperref}

\makeatletter       

\renewcommand{\p@subsection}{}

\makeatother

\newcommand*{\eweakgroup}{\mbox{$SU(2)_L \times U(1)_Y$} }
\newcommand*{\emgroup}{\mbox{$U(1)_{em}$} }

\begin{document}

\title{Light dark matter in a minimal extension with two additional real singlets}

\author{Markos Maniatis\footnote{\url{maniatis8@gmail.com}}}

\affiliation{Centro de Ciencias Exactas and Departamento de Ciencias B\'asicas, UBB,\\
Avda. Andres Bello 720,
Chill\'a{}n, Chile.}


\begin{abstract}
The direct searches for heavy scalar dark matter with a mass of order 100~GeV are much more sensitive than for
light dark matter of order 1~GeV.  The question arises whether dark matter could be light and has escaped detection so far. We study a simple extension of the Standard Model 
with two additional real singlets. 
We show that this simple extension may provide the observed 
relic dark matter density, does neither disturb big-bang nucleosynthesis nor the cosmic microwave 
background radiation observations and fulfills the conditions of clumping behavior for different sizes of 
galaxies. The potential of one Standard Model-like Higgs-boson doublet and the two singlets 
gives rise to a changed Higgs phenomenology, 
in particular, an enhanced invisible Higgs-boson decay rate is expected, detectable by missing transversal momentum searches at the ATLAS and CMS experiments at CERN.
\end{abstract}

\maketitle
\flushbottom



\section{Introduction}
\label{intro}

Recently it has been reported~\cite{vanDokkum:2018vup} 
that the NGC1052--DF2 galaxy with  a stellar mass of approximately \mbox{$2 \times 10^8$} solar masses has a
 rotational  movement in accordance with its observed mass. If this negative indirect search for dark matter~(DM)
 is confirmed it challenges any attempt to solve the puzzle of galaxy rotations by modifications of gravity. 
 
 On the other hand we are facing very severe negative results from direct searches of DM, for instance from
  the cryogenic experiments
  SuperCDMS~\cite{Scorza:2013eto}, CRESST~\cite{CRESST:2017cdd}, EDELWEISS~\cite{Kozlov:2013zuy},  
  and the noble liquid experiments  ArDM~\cite{Calvo:2015uln}, DarkSide~\cite{Wada:2017ftr}, DEAP~\cite{Ajaj:2019imk}, 
  LUX~\cite{daSilva:2017swg}, PandaX~\cite{Xia:2018qgs}, WARP~\cite{Fiorillo:2008zz}, XENON~\cite{Aprile:2019dbj}, and
  ZEPLIN~\cite{Araujo:2012zz}. 
  These experiments provide very low upper limits for the interaction cross sections of DM with nucleons.
  For instance, the XENON100~\cite{Aprile:2012nq} direct detection experiment at Gran Sasso excludes spin-independent elastic 
  nucleon cross sections down to values as tiny as \mbox{$2 \times 10^{-45} \text{cm}^2$} for DM particles with a mass of 55 GeV at 90\% confidence level.

In case that the Higgs boson couples to DM, also collider experiments 
may detect DM; for an overview of this subject we refer to~\cite{Englert:2011yb}.
Even that DM is expected not to show traces in the detector components,
since it is expected to couple only very weakly to 
Standard Model~(SM) particles, it may be discovered as missing transversal momentum in collisions. 
In particular, in this way a DM candidate may enhance the invisible decay rate of
a SM-like Higgs boson. Note that in the SM the only invisible decay channel
of the Higgs boson ($h$) is via two electroweak $Z$ bosons which subsequently
decay into pairs of neutrinos ($\nu$), that is, $h \to ZZ \to 4 \nu$. 

The collider experiments ATLAS~\cite{Aad:2008zzm} and CMS~\cite{Chatrchyan:2008aa}
 have measured upper limits 
on the spin-independent DM-nucleon scattering cross section~\cite{Sirunyan:2018owy,ATLAS:2020cjb}.
Under the assumption that DM couples to the Higgs boson, these measurements provide the most stringent bounds available on light DM detection, far below the underground DM-nucleon direct-detection limits.
However, the upper bounds for light {\em scalar} DM are orders of magnitudes
less stringent than the bounds for heavy DM particles with 
masses of ${\cal O} (100 \text{ GeV})$.

Let us in this context recall that interactions
of DM with SM particles can also not be arbitrarily small: supposing that DM is produced
dynamically in the evolution of the Universe, 
 for decreasing annihilation cross sections of the DM
particles to Standard Model particles, the annihilation processes become more and more rare in the evolution of the Universe and freeze out happens earlier -- corresponding to a higher DM number density. Eventually this would lead to an overclosure of the Universe.

One example of this is the so called Lee-Weinberg bound~\cite{Lee:1977ua}: in case of leptonic dark matter it has been shown that its mass 
has to be greater than of the order of 2~GeV in order not to lead to an overclosure of the Universe. 
However this bound refers to dark matter lepton candidates. If instead the 
annihilation is not mediated by electroweak gauge bosons this bound does not apply -- opening the window for light dark matter.

Supersymmetric models provide generically new weakly interacting bosons and fermions, and therefore in principle these models can provide light DM particles. However, at least the simplest supersymmetric extension of the Standard Model
appears to be disfavored, since there is a tension of the predicted Higgs mass spectrum in contrast to observations --
in particular the detected Higgs boson is too heavy in order to be a supersymmetric particle in the minimal supersymmetric extension (see for instance the review \cite{Maniatis:2009re}).

Here we want to study a simple model with two additional real scalars.
 One of theses scalars serves as a DM candidate ($\chi$) and the other is a scalar 
 {\em mediator}~($\eta$). 
 Since we expect that the DM particles $\chi$ 
 annihilates into a pair of scalar mediators $\eta$, we are not confronted with the Lee-Weinberg bound
and we may have sufficiently large annihilation cross sections $\chi \chi \to \eta \eta$. On the other hand,
through couplings of the mediator $\eta$ to neutrinos, similar to the studies~\cite{Ma:2017xxj, Ma:2016tpf}, 
this annihilation cross section does neither disturb big bang nucleosynthesis nor the observed 
cosmic microwave background radiation~\cite{Galli:2009zc}. 
Moreover, through the coupling of DM~$\chi$ to the mediator $\eta$ we automatically 
get DM self-interactions~\cite{Donato:2009ab, Bringmann:2016din} which do not contradict 
structure formation simulations for different sizes of galaxies~\cite{Tulin:2017ara}.

The potential of the Higgs doublet $h$ and the two real scalars $\chi$, $\eta$ provides 
couplings among these elementary particles. 
These couplings give a changed Higgs phenomenology compared to the SM.
For instance, from the coupling of the Higgs boson $h$ to the DM candidate $\chi$,
 we get an enhanced invisible Higgs-decay rate at colliders,
 since $\chi$ is expected to be stable and escapes detection.
Therefore it is expected to have an enhanced invisible decay rate at the large hadron collider at CERN at the experiments ATLAS and CMS. 
The signature comes from missing transversal momentum with respect to the collision direction in observations at these experiments. 

Scalar DM has been studied, for instance in \cite{Silveira:1985rk,McDonald:1993ex,Burgess:2000yq,Boehm:2003hm,Arcadi:2017kky} and references therein. 
In \cite{Abada:2011qb, Ahriche:2013vqa,Ma:2017xxj,Arhrib:2018eex} models are studied with the scalar DM candidate accompanied by a mediator, with the DM particle mass of the order of 100~GeV.
For a review of light DM we refer to \cite{Knapen:2017xzo}.
Here we focus on the two-real-scalar extension of the SM with one light DM candidate accompanied by a mediator.


\section{The minimal extension with two additional real singlets}
\label{twosinglet}

In the model we propose we have besides one Higgs-boson doublet 
two real singlets,
\begin{equation}
\varphi(x) = \begin{pmatrix} \varphi^+(x) \\ \varphi^0(x) \end{pmatrix},
\qquad \chi(x), 
\qquad \eta(x).
\end{equation}
We employ the convention that the upper component of the doublet is charged.
The doublet gives as usual masses to
the fermions and gauge bosons.
The potential reads
\cite{Abada:2011qb, Ahriche:2013vqa} 
\begin{equation} \label{pot}
\begin{split}
V_{\text{light}}^{\text{DM}} =&
\mu_h^2 \varphi^\dagger \varphi
+ \mu_\chi^2 \chi^2
+ \mu_\eta^2 \eta^2
+ \lambda_h (\varphi^\dagger \varphi)^2
+ \lambda_\chi \chi^4
+ \lambda_\eta \eta^4\\
&
+ \lambda_{h \chi} (\varphi^\dagger \varphi) \chi^2
+ \lambda_{h \eta} (\varphi^\dagger \varphi) \eta^2
+ \lambda_{\chi \eta} \chi^2  \eta^2 \;,
\end{split}
\end{equation}
where we suppress the argument of the fields from here on in most cases.
Inspecting this potential we encounter apart from the electroweak symmetry $\eweakgroup$ 
the discrete symmetries  $\mathbbm{Z}_2$ with $\chi$ transforming as $\chi \stackrel{\mathbbm{Z}_2}{\longrightarrow} -\chi$
and $\mathbbm{Z}_2'$ as $\eta \stackrel{\mathbbm{Z}_2'}{\longrightarrow} -\eta$, where we assume that the Higgs doublet
as well as all Standard Model fields transform trivially.
We want the potential to provide vacuum expectation values for the neutral component of the doublet, $\varphi^0$,
as well as for the field $\eta$, that is, besides electroweak symmetry breaking
$\eweakgroup \to \emgroup$,
the potential should break $\mathbbm{Z}_2'$ spontaneously.
Since the symmetry $\mathbbm{Z}_2$ is not broken, neither explicitly nor spontaneously, we expect a 
stable particle~$\chi$ providing DM.

The domain wall problem~\cite{Zeldovich:1974uw}, appearing inevitably for a 
spontaneously broken discrete symmetry, is expected to be circumvented in the usual way: we assume that the discrete symmetry $\mathbbm{Z}_2'$ is broken explicitly by an additional cubic mass-parameter term linear in $\eta$. This additional term however is Planck-mass suppressed and therefore only of relevance at very high energies and therefore negligible in our analysis here and not further taken into account.

We want the mediator $\eta$ to decay sufficiently fast in order to not disturb big-bang nucleosynthesis. 
This can be achieved by a coupling of $\eta$ to right-handed neutrinos. We arrange theses couplings by
imposing the right-handed neutrinos to transform under $\mathbbm{Z}_2'$ as
\begin{equation} \label{nRtrans}
\nu_{R, e} \stackrel{\mathbbm{Z}_2'}{\longrightarrow} +\nu_{R, e}, \quad
\nu_{R, \mu} \stackrel{\mathbbm{Z}_2'}{\longrightarrow} -\nu_{R, \mu}, \quad
\nu_{R, \tau} \stackrel{\mathbbm{Z}_2'}{\longrightarrow} -\nu_{R, \tau},
\end{equation}
and add appropriate Majorana kinetic terms and mass terms
\begin{equation} \label{majorana}
{\cal L}_{\nu_R} = i \bar{\nu}_{R, i} \slashed{\partial}  \nu_{R, i}
- \left( \frac{\lambda_{i j}}{2}\; \eta \; {\bar{\nu}_{R, i}}^c \nu_{R, j} + h. c. \right) \;,
\end{equation}
with $i, j  \in \{ e, \mu ,\tau\}$ the indices denoting the three flavors of the neutrinos. 
Different choices for the $Z_2'$ symmetry behavior of the right-handed neutrinos in~\eqref{nRtrans} could be chosen, however this assignment keeps
the number of parameters for the matrix $\lambda_{ij}$ rather restricted.

The parameters should provide stability of the potential and the observed electroweak symmetry breaking, that is,
in this case we have a global minimum of the potential with
\begin{equation}
\langle \varphi \rangle = \begin{pmatrix} 0  \\  \frac{1}{\sqrt{2}} v_h  \end{pmatrix},
\qquad
\langle \eta \rangle = v_\eta,
\qquad
\langle \chi \rangle = 0.
\end{equation}
We write the neutral component of the Higgs-boson doublet and 
the scalar $\eta$ expanded about their respective vacuum-expectation values,
\begin{equation}
\varphi(x) = \begin{pmatrix}  \varphi^+(x)  \\  \frac{1}{\sqrt{2}} \left(v_h + h(x)\right)  \end{pmatrix}, 
\qquad
\eta(x) = v_\eta + \eta_e(x),
\end{equation}
with $h(x)$ and $\eta_e(x)$ real fields.
The necessary tadpole conditions for stability at the vacuum yield 
 \begin{equation} \label{tad}
  \frac{1}{2} v_h^2 = \frac{ \lambda_{h \eta} \mu_\eta^2 - 2\lambda_\eta \mu_h^2}{ 4 \lambda_h \lambda_\eta - \lambda_{h \eta}^2}, \quad
 \frac{1}{2} v_\eta^2 =  \frac{ \lambda_{h \eta} \mu_h^2 - 2\lambda_h \mu_\eta^2}{ 4 \lambda_h \lambda_\eta - \lambda_{h \eta}^2} . 
\end{equation}

For the DM mass squared we find from the potential, that is, from the quadratic terms with respect to the DM field $\chi$ after electroweak symmetry breaking,
\begin{equation}
m_\chi^2 = 2 \mu_\chi^2 + \lambda_{h \chi} v_h^2 + 2 \lambda_{\chi \eta} v_\eta^2\;.
\end{equation}

We get for the mixing matrix of the excitations $h$ and $\eta_e$ 
in the  basis $( h, \eta_e)$  
 \begin{equation} \label{hetamix}
 \begin{pmatrix}
 2 v_h^2 \lambda_h & 
 v_h v_\eta \lambda_{h \eta} \\
 v_h v_\eta \lambda_{h \eta}  &   2 v_\eta^2 \lambda_\eta.
 \end{pmatrix}.
 \end{equation}
This mixing matrix is diagonalized by an orthogonal matrix with mixing angle $\theta$,
 \begin{equation} \label{theta}
 \tan ( 2 \theta) = 
 \frac{v_h v_\eta \lambda_{h \eta} }
 {v_h^2 \lambda_h -  v_\eta^2 \lambda_\eta}
 \end{equation}
and we get the mass eigenstates $h'$ and $\eta'$.
In order to get a Higgs boson with the observed mass of about~125~GeV and a light mediator, we see that the mixing angle, or equivalently the parameter $\lambda_{h \eta}$, has to be rather tiny.

Eventually, from the spontaneous breaking with respect to $\eta$,
 the right-handed neutrino-mass matrix is generated,
 \begin{equation} \label{Mneutrino}
 (M_\nu)_{ij} = \lambda_{ij} \langle \eta \rangle, \quad i,j = e , \mu, \tau \;.
 \end{equation}
 

\section{Phenomenology of dark matter and the mediator}

Let us start with a study of the relic abundance of DM in the 
model with two additional real scalars.
With view on the potential~\eqref{pot} we see that DM ($\chi$) may annihilate
into pairs of mediators $\eta'$ (the mass eigenstate is denoted by $\eta'$).
This annihilation arises from the quartic $\chi$-$\chi$-$\eta$-$\eta$ coupling, 
as well as via the trilinear $\chi$-$\chi$-$h$ and $\chi$-$\chi$-$\eta$ couplings which come from the quartic couplings after spontaneous symmetry breaking. 
At tree level we get from these interactions besides the direct coupling two $s$-channel diagrams with a Higgs-boson, respectively mediator propagator and one $t$-channel diagram with a mediator propagator. The $s$-channel diagram with a Higgs-boson propagator is kinematically suppressed due to the light DM and mediator masses compared to the Higgs-boson mass. Therefore the relic density of DM
depends mainly on the 
coupling parameter $\lambda_{\chi \eta}$ besides
 the masses $m_\chi$ and $m_{\eta'}$.
For $\eta'$ lighter than $\chi$, the DM annihilation cross section is 
too large, freeze out occurs too late and the resulting relic density is too small.
\begin{figure*}[th!]
\centerline{
\includegraphics[width=1.0\textwidth, angle=0]{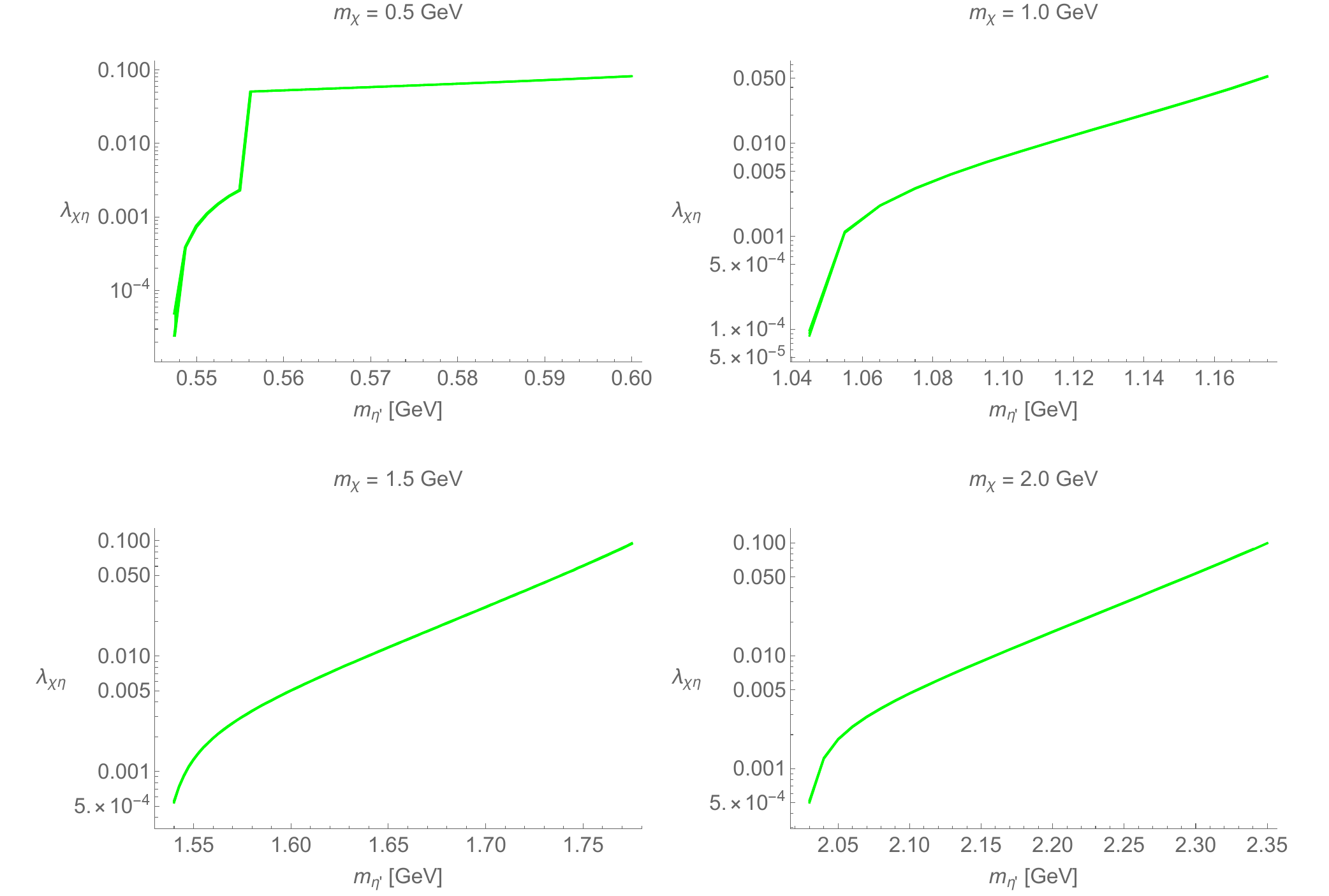}
}
\caption{\label{relic} 
Mass of the mediator $m_{\eta'}$ versus the quartic coupling parameter~$\lambda_{\chi \eta}$
providing the observed relic density of $\Omega_{\text{DM}} h^2 = 0.1200(12)$~\cite{Aghanim:2018eyx} (with 1-$\sigma$ uncertainty). 
The study shows the values for four different choices of a light DM particle, that is,  $m_\chi$ equals 0.5~GeV
(upper left), 1.0~GeV (upper right), 1.5~GeV (lower left), and 2.0~GeV (lower right). The width of the line
corresponds to the 1-$\sigma$ variation.
}
\end{figure*}

In Fig.~\ref{relic} we show the values for the mediator mass $m_{\eta'}$ versus the quartic parameter~$\lambda_{\chi \eta}$
which provide the observed relic density of $\Omega_{\text{DM}} h^2 = 0.1200(12)$~\cite{Aghanim:2018eyx} (the value in parenthesis gives the 1-$\sigma$ uncertainty). 
This study is done for four different values of the DM mass,
that is, $m_\chi$ in the range from 0.5~to 2~GeV as indicated in the figure.
From this study we see that it is possible for a light DM candidate accompanied by a
slightly heavier mediator to achieve the right, that is, indirectly observed relic density. 
We see that for increasing values of the mediator mass $m_{\eta'}$, the right
relic density can be obtained by strongly raising the
quartic coupling~$\lambda_{\chi \eta}$ over orders of magnitude as shown in this semi-logarithmic figures. 
As mentioned above, the annihilation of dark matter into mediators occurs besides the quartic DM mediator coupling via $s$- and $t$-channel diagrams. 
Varying the
mediator mass we get competing contributions of the remaining different diagrams corresponding to the different regions of the mediator mass. This behavior is reflected by 
the kink structure in Fig.~\ref{relic}.

As discussed later in the context of DM self-interactions, we want to avoid a too strong 
DM self-interaction, therefore we set the 
quartic parameter $\lambda_\chi$ to zero and consider rather small values for the parameter~$\lambda_{\chi \eta}$. 
The relic density is numerically calculated from the solutions of the Boltzmann equation 
using the package MICROMEGAS~\cite{Belanger:2013oya, Belyaev:2012qa}.
All other parameters are fixed to $m_{h'}=125 \text{ GeV}$,
$v_h = 246 \text{ GeV}$ (the standard electroweak parameters), $v_\eta = 3 \text{ GeV}$,
$\lambda_{h \eta} = 1 \times 10^{-7}$, $\lambda_{h \chi} = 0.1$,
 a vanishing parameter $\lambda_{\chi}$,
 and we fix 
the  non-vanishing right-handed neutrino-mixing-matrix parameters \eqref{Mneutrino} 
to $\lambda_{e \mu} = \lambda_{e \tau} = 0.1$.
The remaining parameters $\lambda_h$, $\lambda_\eta$ and the sine of the Higgs-mediator mixing angle,
$\sin{\theta}$, are then determined by the tadpole conditions~\eqref{tad}
 
We proceed with a study of DM self-interactions; for an introduction to this subject we refer to~\cite{Tulin:2017ara}. 
Observations show rather flat distributions in the cores in smaller galaxies of dwarf size 
up to galaxies of the size of our Milky Way. 
In contrast, simulations of galaxy formations,
assuming that DM is collisionless,
predict a cusp in the density profile for these sizes of galaxies.
This mismatch between observation and simulation can be resolved if DM 
has self-interactions 
 $\chi \chi \to \chi \chi$  
with a cross section per DM mass of the order of $\sigma/m_\chi \approx 1 \text{ cm}^2/g$~\cite{Tulin:2017ara}
for smaller galaxy sizes up to our Milky Way and is approximately collisionless for
sizes larger than our Milky way up to galaxy clusters, that is, with self-interaction cross sections per DM mass
dropping to values below $0.1 \text{ cm}^2/g$.
Typical rotational velocities for dwarf galaxies are 
10~km/s, for Milky Way type galaxies 200~km/s and for galaxy clusters 1000~km/s~\cite{Tulin:2017ara}.

In the model considered here with two additional scalars, 
the self-interactions arise, as can be seen from the potential, 
from the quartic DM self coupling with parameter $\lambda_\chi$, from
the $\chi$-$\chi$-$\eta$-$\eta$ coupling with parameter $\lambda_{\chi \eta}$,
and from the $\chi$-$\chi$-$h$-$h$ coupling with parameter $\lambda_{h \chi}$.
Since we are considering light DM, the contribution from the SM-like Higgs boson $h'$
is suppressed. The quartic coupling $\lambda_\chi$ on the other hand has
to be suppressed also since otherwise we immediately overshoot the self-interaction
cross sections.
In addition to the quartic $\chi$ interaction, we encounter self-interactions from trilinear couplings
of $\chi$ with the Higgs boson $h$ and the mediator $\eta$ in $s$- and $t$-channel diagrams. 
\begin{figure*}[th!]
\includegraphics[width=1.0\textwidth]{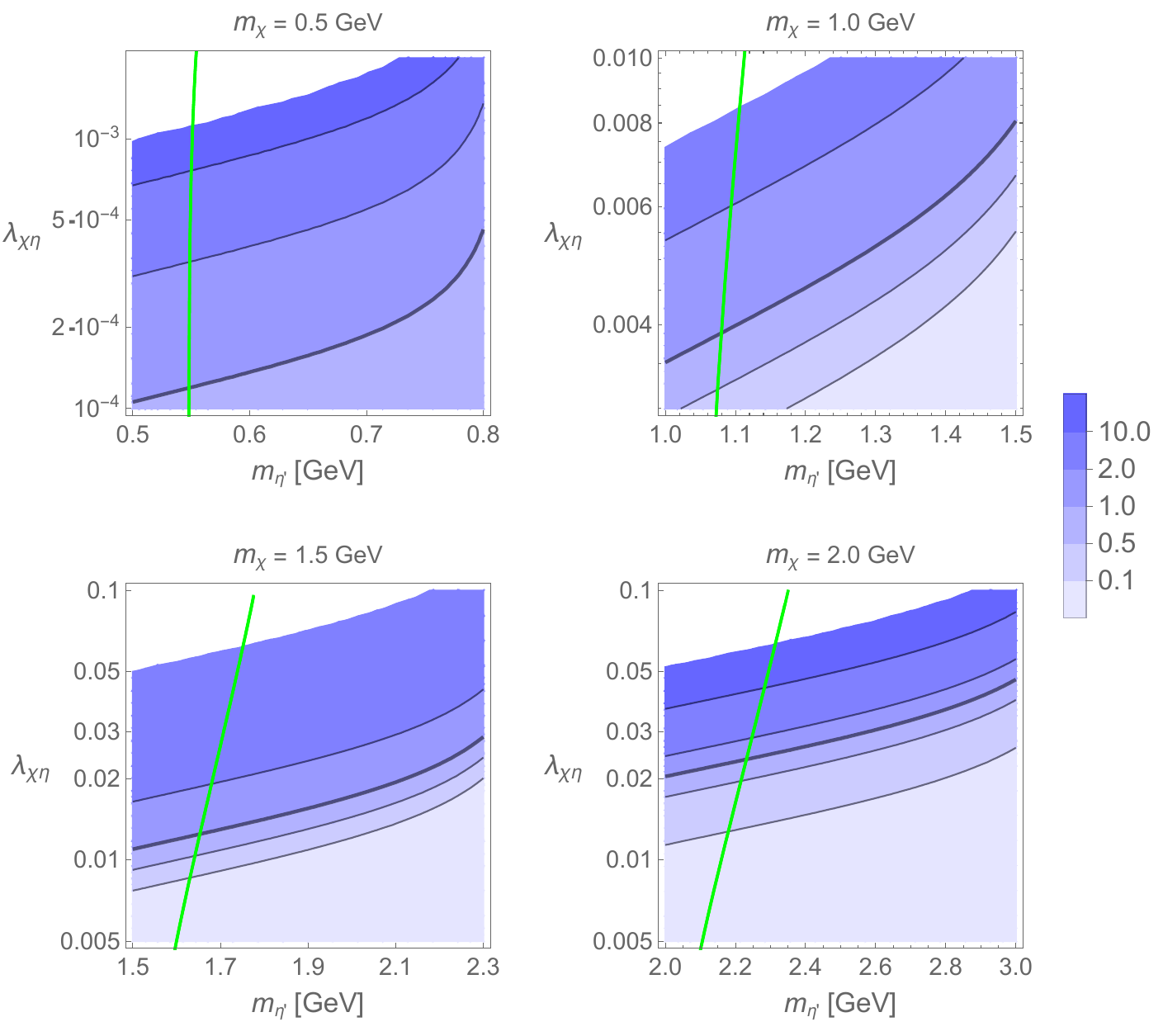}
\caption{\label{selchiRD}
Contour plots of self-interacting DM cross section per DM mass $\sigma(\chi \chi \to \chi \chi)/m_\chi$
in the $m_{\eta'}$-$\lambda_{\chi \eta}$ plane for four different values of the light DM mass, that is,
$m_\chi$ equals 0.5~GeV
(upper left), 1.0~GeV (upper right), 1.5~GeV (lower left), 2.0~GeV (lower right). The legend 
gives the regions of different values of the DM self-interaction per DM mass in units of $\text{ cm}^2/\text{g}$, 
the central black curve in the figures corresponds to $\sigma(\chi \chi \to \chi \chi)/m_\chi =1 \text{ cm}^2/\text{g}$.
In the calculation of the cross section per DM mass, the relative velocity of DM particles is set to 10~km/s corresponding to dwarf galaxies.
The steeply raising 
green line replicates the results from Fig.~\ref{relic} corresponding to the right DM relic density of 
$\Omega_{\text{DM}} h^2 = 0.1200(12)$~\cite{Aghanim:2018eyx} (with 1-$\sigma$ uncertainty). 
  }
\end{figure*}

In Fig.~\ref{selchiRD} we show a study of DM self-interactions. The contour plots 
show the mediator mass $m_{\eta'}$ versus 
the quartic coupling parameter $\lambda_{\chi \eta}$ with respect to the DM self-interaction 
cross section per DM mass. Similar to the study of the DM relic density we have varied the value
of DM in the range of 0.5~GeV to 2.0~GeV as indicated in the figure. 
We have chosen a relative velocity of 10~km/s, corresponding to dwarf galaxies with 
a required value of $\sigma(\chi \chi \to \chi \chi)/m_\chi =1 \text{ cm}^2/\text{g}$~\cite{Tulin:2017ara}.
We see that it is possible to get the observed value for the self-interaction. 
In this figure we also replicate the results from the DM relic density study, as shown previously 
in Fig.~\ref{relic}. The correct relic density curve (green line) raises much steeper than the contours for the self-interaction cross section per DM mass (black line) with respect to increasing values of the mediator mass~$m_{\eta'}$. 
All other parameters are set as above in the study of the DM relic density. 
In particular we see that both curves intersect, giving parameter values which on the one hand 
provide the observed DM relic density and on the other hand provide the correct 
DM-self-interaction cross section.
As we can see in this study, even for a vanishing quartic DM-coupling parameter 
$\lambda_\chi$ the quartic $\chi$-$\chi$-$\eta$-$\eta$ parameter $\lambda_{\chi \eta}$ has to be rather small.
In principle multiple exchanges between the DM particles could lead to Sommerfeld enhancement~\cite{Feng:2010zp} of the self-interactions, however these long-range interactions are for the here considered similar masses of DM and the mediator negligible. Let us note that the Sommerfeld enhancement is implemented in the MICROMEGAS package.

We study also the 
dependence of the self-interaction cross section per DM mass on the 
relative velocity of DM.
 From the explicit calculation of the cross section we find that the self-interacting cross section
 per DM mass drops orders of magnitude with increasing relative velocities. 
 In particular,  if we arrange the parameters such that we get for a relative velocity of 
 $v_{\text rel}=10 \text{ km/s}$ (corresponding to dwarf galaxies) the wanted cross section per DM mass
 of $\sigma(\chi \chi \to \chi \chi)/m_\chi =1 \text{ cm}^2/\text{g}$, then we 
 find for larger values of the relative velocity, that is, $v_{\text rel} > 200 \text{ km/s}$ (corresponding 
 to galaxy sizes larger than our Milky Way) values below $0.1 \text{ cm}^2/\text{g}$. 
 Therefore, the study as shown in Fig.~\ref{selchiRD} gives
 parameters which not only give the right self-interactions of DM, but also
 naturally drop to values in agreement with observations of density profiles of 
 different sizes of galaxies.

 In the early Universe the mediators thermalize with the right-handed neutrinos, but below a temperature of about the mass of the mediator, that is, of the GeV order, the right-handed neutrinos decouple. 
 Since they are stable they contribute to the relic neutrino density. 
The energy density of radiation of neutrinos~$\rho_\nu$ with respect to
photons~$\rho_\gamma$ after the time of electron-positron annihilation is~\cite{Zyla:2020zbs}
\begin{equation}
\frac{\rho_\nu}{\rho_\gamma} 
=
\frac{7}{8} 
 N_{\text{eff}}^\nu 
\left( \frac{4}{11} \right)^{4/3},
\end{equation}
where the fraction $7/8$ originates from the fermionic Nature of neutrinos and the factor $(4/11)^{1/3}$ comes from the reheating of the 
photons from electron-positron annihilation.
In the Standard Model, the number of effective neutrinos has been calculated to 
$N_{\text{eff}}^{\text{SM}} = 3.045$~\cite{deSalas:2016ztq} for three neutrino
species. Every additional neutrino contributes to the number of effective neutrinos~\cite{Steigman:1979xp,Zhang:2015wua},
\begin{equation}
\Delta N_{\text{eff}} = 
\left( \frac{ g(T^L_{\text{dec}}) } {g(T^R_{\text{dec}})} \right)^{4/3}
\end{equation}
with $g(T)$ the effective number of degrees of freedom at temperature $T$, where the respective neutrinos decouple.
Since the left-handed neutrinos decouple at the MeV scale, but the right-handed neutrinos in the model considered here decouple at the GeV scale, we find from~\cite{Husdal:2016haj} with  
$g(MeV) = 9.5$ and $g(GeV)= 76.5$ 
for every right-handed neutrino the rather small contribution, 
$\Delta N_{\text{eff}} = 0.062$. 
On the other hand we get from the recent cosmic microwave background measurement from the Planck satellite experiment (combined with baryon acoustic oscillation),
$N_{\text{eff}} = 2.99 \pm 0.17$~\cite{Aghanim:2018eyx}. 
This means that 
the measurement of $N_{\text{eff}}$  is in agreement with the light DM model considered here. 
We conclude that 
 no substancial change
of microwave background radiation, nor big-bang nucleosynthesis is to be expected (for the case of new scalars coupled to Standard Model particles we refer to 
the overview~\cite{Clarke:2013aya}).

Eventually we want to study the Higgs phenomenology in the model considered here. 
Compared to the SM, the Higgs-boson doublet is accompanied by two additional scalars, $\chi$ and $\eta$. 
The potential~\eqref{pot} involves in addition to the quadratic and quartic 
self couplings also couplings among the three scalars.
As discussed in section~\ref{twosinglet} we encounter a 
mixing of the neutral component of the doublet $h$ with the mediator $\eta$ forming 
a Higgs mass eigenstate $h'$.
From~\eqref{theta} we see that the tangent of twice the mixing angle is proportional to the potential parameter $\lambda_{h \eta}$. 

Since the Higgs boson $h$ couples on the one hand  
 to the DM particle $\chi$  
 with coupling strength $\lambda_{h \chi}$
 and on the other hand has 
 the usual Yukawa couplings to the fermions, we see
that DM interacts with the hadrons of the nucleon.
This DM-nucleon interaction is therefore expected to be
detectable at direct detection experiments -- in the introduction 
we have mentioned some direct-detection experiments searching 
for DM-nucleon interactions.

\begin{figure*}[th!]
\centerline{
\includegraphics[width=0.6\textwidth, angle=0]{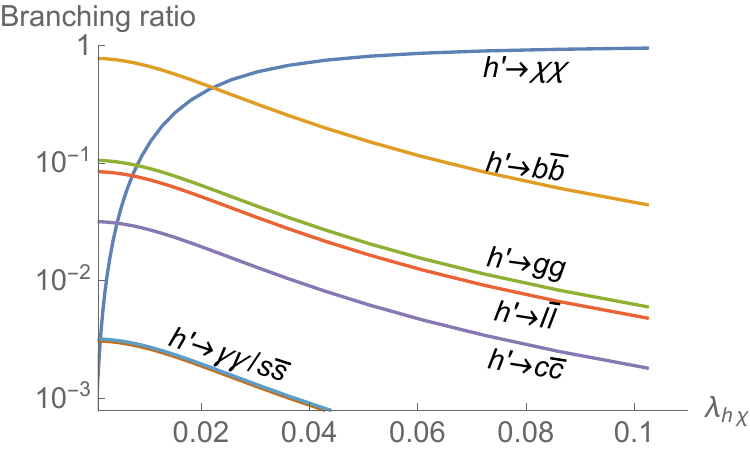}
}
\caption{\label{hpbranch} 
Branching ratios of the SM-like Higgs boson~$h'$ depending on the quartic potential 
parameter~$\lambda_{h\chi}$.
The decay rates are for a pair of DM particles $\chi\chi$,  for a $b$-quark pair $b\bar{b}$, gluons $gg$, leptons $l\bar{l}$, 
$c$-quarks $c\bar{c}$, and branching rates of nearly the same size for photons $\gamma \gamma$ and a $s$-quark pair~$s\bar{s}$.
}
\end{figure*}

Moreover, the coupling of the DM candidate to the Higgs boson opens the
possibility to detect DM at collider experiments.
In particular, since we are considering light DM, that is, we assume to have $m_{h'} > 2 m_{\chi}$, 
the SM-like Higgs boson $h'$ can decay into a pair of DM particles $\chi$.
Since the DM particles are assumed to be stable, these decays do not show
visible traces in the detector components of collider experiments.
However, these events can be discovered from the observation of missing transversal momentum.
The invisible decay width of the SM-like Higgs boson $h'$ into a pair of $\chi$'s 
at tree-level reads
\begin{equation} \label{lhinv}
\Gamma^{\text{inv}}_{h' \to \chi \chi} =
\frac{\lambda_{h \chi}^2 v_h^2 \beta_\chi}{8 \pi m_{h'}}, \quad \text{with } \beta_\chi = \sqrt{1 - 4 \frac{m_\chi^2}{m_{h'}^2}}.
\end{equation}
This decay width is obviously proportional to the quartic potential parameter 
$\lambda_{h \chi}$ squared.
In Fig.~\ref{hpbranch} we 
study the branching ratios of the SM-like Higgs boson $h'$ varying 
this quartic coupling parameter $\lambda_{h\chi}$.
In this study we have chosen all other parameters as mentioned above
with a DM mass of $m_\chi = 0.5$~GeV and the mediator slightly heavier, that is, $m_{\eta'} = 0.55$~GeV.  This pair of values together with the parameter $\lambda_{\chi \eta}= 10^{-4}$ yields the correct values for the relic density of DM and for the DM self-interactions; see Fig.~\ref{selchiRD}.
Since the Higgs boson~$h'$ has a tiny mediator~$\eta$ component it could also decay into neutrinos and in addition there are couplings of the Higgs boson to mediator particles. 
However for the tiny mixing parameter $\lambda_{h \eta}$ considered here these contributions are negligible, as shown explicitly in~Fig.~\ref{hpbranch}.

As we can see in this study we find for larger values of the parameter~$\lambda_{h \chi}$ a dominant Higgs-decay channel into DM particles $\chi$. 
Depending on the parameter $\lambda_{h\chi}$ we find therefore an enhanced invisible
branching ratio compared to the SM. Let us remind us that in the SM the Higgs boson can only decay invisibly 
via $Z$ bosons which subsequently decay into neutrinos, that is, $h \to ZZ \to 4 \nu$.

Since the invisible decay rate it restricted from the missing transversal momentum searches at the CMS and ATLAS experiments at the LHC this translates into
an upper bound on the coupling strength between the Higgs boson and the DM particle.
The ATLAS collaboration has recently measured~\cite{ATLAS:2020cjb} the upper limits of the invisible branching ratio and finds 0.13 at the 95\% confidence level in agreement
with the Standard Model. The CMS collaboration has published~\cite{Sirunyan:2018owy} an invisible branching ratio of 0.24 compared to the Standard Model expectation of 0.23.
From the calculation of the branching ratios as shown in Fig.~\ref{hpbranch} we get for an upper bound of 0.13 for the invisible branching ratio the constraint $\lambda_{h \chi} < 0.01$
for the potential parameter.
Let us also mention the context of the coupling parameter $\lambda_{h \chi}$
to the spin-independent DM-nucleon cross section, which reads~\cite{Djouadi:2011aa}
\begin{equation} \label{DMnucl}
\sigma_{\chi N}^{SI} =
\frac{\lambda_{h \chi}^2}{\pi m_h} 
\frac{m_N^4 f_N^2}{(m_\chi + m_N)^2}.
\end{equation}
Here $m_N$ is the nucleon mass and $f_N$ a nucleon form factor.
Based on phenomenological and lattice-QCD calculations, a Higgs-nucleon
form factor of $f_N = 0.308(18)$ has been calculated~\cite{Hoferichter:2017olk}.
The coupling parameter $\lambda_{h\chi}$ which appears in the DM-nucleon 
cross section~\eqref{DMnucl} can now, using \eqref{lhinv}, be replaced by
the invisible decay rate. Therefore, the measured upper bound on the invisible decay rate in Higgs-boson decays
translates into an upper bound for the DM-nucleon cross section~\cite{Djouadi:2011aa}.
The 90\% upper confidence limits on the spin-independent DM-nucleon scattering cross section are compared to the direct detection limits in~\cite{Sirunyan:2018owy}. 
It is shown that the CMS and ATLAS results from the measurement of the
invisible decay of the Higgs boson are orders of magnitudes more sensitive than
direct detection experiments for scalar DM masses up to 7~GeV. These are currently the most stringent limits available -- assuming that there is a Higgs-DM coupling.

\section{Conclusions}

The rotational movement of stars in galaxies is typically faster than expected from the visible mass. 
Attempts to modify gravity appear disfavored due to the observation of exceptional galaxies showing no mismatch.
New, yet undiscovered, elementary dark matter particles could solve the puzzle.
However, direct detection experiments of dark matter-nucleon interactions 
have not revealed any dark matter particle.
The lowest bounds in these experiments come from 
searches of dark matter particles with a mass 
of the order of 100~GeV.
Therefore it appears quite natural to look for light DM particles.
In case the light DM candidate is leptonic with a mass below 2~GeV 
-- the so-called Lee-Weinberg bound --
this would lead to an overclosure of the Universe.
Here we have discussed a simple model with two additional scalars, odd under discrete symmetries, not restricted by the Lee-Weinberg bound.
The scalar potential of the model provides interactions among the DM candidate, a mediator and a SM-like Higgs boson. These interactions provide DM annihilation into a pair of mediators.
The mediators in turn decay into pairs of right-handed neutrinos. This process is sufficiently fast, that is, before big-bang nucleosynthesis
takes place, and in particular neither DM nor the mediator decay into Standard Model particles. Therefore we neither expect displaced vertices in colliders, decays into muons, nor changed meson decays.

The objective of this study is to investigate the agreement of the simple light DM model with different observations in telescopes, underground experiments, and 
collider experiments with respect to DM.
The relic density in this two-real-scalar extension has been computed, and it has been shown that, for a dark matter 
mass in the range of 0.5~GeV to 2~GeV and a 
mediator slightly heavier, we can get the observed relic density. 
Also we have studied dark matter self-interactions which may explain the observation of rather flat DM density profiles in cores of dwarf galaxies up to galaxies of the size of our Milky Way. In the model considered here the interaction of DM with the mediator provides these self-interactions quite naturally.
Moreover, due to the Nature of these interactions transmitted by the mediator, 
these self-interactions automatically drop for larger relative velocities in galaxy Clusters -- in agreement with simulations of galaxy formations.

Since the mediator decays into right-handed neutrinos which are
stable, these neutrinos contribute
to the neutrino background radiation. Indirectly this 
background could change big-bang nucleosynthesis or
the cosmic microwave background. This effect is encoded in the number of effective
neutrinos which is measured at the Planck satellite experiment
yielding 
$N_{\text{eff}} = 2.99 \pm 0.17$~\cite{Aghanim:2018eyx}, whereas
in the SM the calculation gives $N_{\text{eff}}^{\text{SM}} = 3.045$~\cite{deSalas:2016ztq} in agreement with the one sigma
deviation of the observation. In the two-singlet model we 
calculate that every right-handed neutrino contributes 
$\Delta N_{\text{eff}} = 0.062$ to the effective number of neutrinos.
This originates from the fact that the right-handed neutrinos decouple
earlier as the left-handed neutrinos in the evolution of the Universe.
The effective number of neutrinos is therefore in agreement with 
observation.

Eventually we have studied the Higgs-boson phenomenology compared to the Standard Model. The interactions of the Higgs boson with the DM candidate and the mediator
give on the one hand a mixing of the Higgs boson with
the mediator and on the other hand provides interactions of the 
Higgs boson with the DM particle. These latter interactions
in turn give rise to interactions of DM with nucleons -- observable at underground DM-nucleon detection experiments,
but also detectable at collider experiments: the Higgs bosons produced at LHC in
the CMS or ATLAS detectors may decay into stable dark matter particles, detectable as events with missing transverse momentum. 
Upper bounds on such searches 
have been published by both experiments and are here translated into upper bounds on the corresponding DM-Higgs coupling parameter. 

Eventually we would like to mention that it is quite striking that a simple extension of the Standard Model with two real singlets 
complies with the different types of DM searches discussed and also shows no contradiction with 
our understanding of the evolution of the Universe.


\begin{acknowledgments}
The work is supported, in part, by the Grants UBB  {\em Materia obscura y los bosones de Higgs}, No. DIUBB 193209~1/R and FONDECYT regular with No. 1200641.
\end{acknowledgments}


\end{document}